\documentclass[%
 reprint,
superscriptaddress,
 amsmath,amssymb,
 aps,
]{revtex4-2}

\usepackage{graphicx}
\usepackage{dcolumn}
\usepackage{bm}
\usepackage{braket}
\usepackage[colorlinks,linkcolor=blue,citecolor=blue]{hyperref}
\usepackage{enumitem}
\usepackage{orcidlink}
\usepackage{dsfont}

\usepackage{booktabs}
\usepackage{float}

\usepackage{tikz}
\usetikzlibrary{decorations.text}
\usetikzlibrary{decorations}
\usetikzlibrary{%
    decorations.pathreplacing,%
    decorations.pathmorphing%
}
\usepackage[ruled, lined, linesnumbered, commentsnumbered, longend]{algorithm2e}

\newcommand{\correction}[1]{\textcolor{black}{#1}}

\begin{document}

\title{Reconstructing the unitary part of a noisy quantum channel}

\author{Adrian Romer}%
\affiliation{%
Freie Universit\"{a}t Berlin, Fachbereich Physik and Dahlem Center for Complex Quantum Systems, Arnimallee 14, 14195 Berlin, Germany
}%
\author{Daniel M. Reich\,\orcidlink{0000-0003-1381-6300}}
\affiliation{%
Freie Universit\"{a}t Berlin, Fachbereich Physik and Dahlem Center for Complex Quantum Systems, Arnimallee 14, 14195 Berlin, Germany
}%
\author{Christiane P. Koch\,\orcidlink{0000-0001-6285-5766}}
\affiliation{%
Freie Universit\"{a}t Berlin, Fachbereich Physik and Dahlem Center for Complex Quantum Systems, Arnimallee 14, 14195 Berlin, Germany
}%

\date{\today}

\begin{abstract}
We consider the problem of reconstructing, from a set of input and output states, the unitary describing the evolution of a quantum system or quantum channel. For ideal, fully coherent evolution, we show that the unitary can be reconstructed from two mixed states or $d+1$ pure states, where $d$ is the size of Hilbert space. The reconstruction method can be extended to approximate the unitary part of a dynamical map, provided the decoherence is not too strong to render this question meaningless. We exemplify the method for the example of the cross-resonance gate as well as a random set of unitaries, comparing the reconstruction from pure, respectively mixed, states to an approach based on the Choi matrix. We find that the pure state reconstruction requires the least amount of resources when the dynamics is close to unitary, whereas the mixed state approach outperforms the pure state reconstruction in terms of channel uses for appreciable decoherence. These conclusions hold also in the presence of  SPAM errors and irrespective of the Hilbert space size.
\end{abstract}

\maketitle
 
\section{Introduction}

The ability to characterize and verify quantum devices is a key requirement for advancing quantum information~\cite{nielsen2010quantum} from basic science to operational technologies.  
Protocols for quantum characterization, verification and validation~\cite{blumekohout2025arXiv} probe the effects of unwanted decoherence so that it can be eliminated or mitigated~\cite{Cai2023RMP-QuErrorMitigation}. At the most comprehensive level, quantum process tomography~\cite{nielsen2010quantum} allows one to 
assess, in a quantitative manner, what is happening inside a quantum device. Unfortunately, its expensive scaling in the number of qubits $N$ prevents its use in practice, as the size of quantum devices increases. More efficient tools have emerged over the past two decades~\cite{blumekohout2025arXiv}, trading the amount of resources that are required for the amount of information that can be obtained with a given method. For example, the popular technique of randomized benchmarking~\cite{blumekohout2025arXiv} allows for estimating the average gate fidelity, i.e., for measuring how well the implemented gate fits the desired one. It does not reveal any information about what evolution has actually been realized. 
Here we focus on this latter question and ask how to reconstruct, at least approximately, the unitary part of a, possibly noisy, quantum channel.  The straightforward approach is to employ quantum process tomography to obtain this answer, since it fully characterizes the complete evolution, both the unitary and the dissipative part. But what is the most efficient approach, when one wants to know the unitary part only?

Combining techniques from compressed sensing with randomized benchmarking, it is possible to recover unitary quantum channels in the multiqubit case ($d = 2^N$) via measurement of gate fidelities of random Clifford unitaries~\cite{roth2018PRL}. This approach involves the solution of a convex optimization problem over the set of unital quantum channels and can be extended to obtain approximations even for non-unitary quantum channels. In this case, the error between the minimiser of the optimisation problem and the unknown channel can be estimated by the measurement noise combined with a measure on the deviation of the channel from unitarity, given by the distance to the best unit rank approximation of the corresponding Choi matrix~\cite{roth2018PRL}.

Here, we propose an alternative approach. 
The setup we envision is as follows: Considering an unknown or badly characterized quantum device, which input states should be prepared, sent through the channel, and measured on output, in order to minimize the number of channel uses? For an ideal, i.e., unitary, channel, the task of identifying the unitary can be thought of as two subtasks --- first, to identify the basis in which the unitary is diagonal and then to determine the eigenvalues via their phases. The minimum number of states is therefore at least two, one that is built up of one-dimensional projectors, weighted with mutually disjoint eigenvalues, allowing to assess the information on the basis, and a second one for the phases, that is rotated with respect to all the one-dimensional projectors~\cite{reich2013PRA}. The second state can be thought of as a basis state taken from a mutually unbiased basis. The two states can be used to estimate the average fidelity of a quantum gate~\cite{reich2013PRA, reich2013prl}, where one compares to a known (and desired) unitary evolution.
Here we consider the more challenging task of reconstructing an unknown operation and 
show that measuring these two states upon output allows 
to reconstruct the unitary part of an unknown channel, provided that the decoherence is not too strong. 
The most obvious application is in channel learning. It is also useful when implementing quantum gates in architectures with limited connectivity where a series of auxiliary operations connecting the targeted qubits may distort the desired evolution in a non-trivial way. Identifying which unitary has been implemented instead of just knowing the error of the desired unitary will allow for designing tailored error  mitigation strategies~\cite{Cai2023RMP-QuErrorMitigation}.

Since mixed states are hard to prepare, we also discuss a variant employing $d$ one-dimensional projectors
instead of encoding  
the basis via one mixed state. While this ensures that all input states are pure, it takes their number from two (irrespective of Hilbert space dimension) to $d+1$, with $d$ the Hilbert space dimension. To benchmark the two state-based reconstruction approaches that we propose, we compare to reconstruction using the Choi matrix, i.e., full quantum process tomography. 

The paper is organized as follows: In Sec.~\ref{sec:approaches_udm} we present our state-based reconstruction approaches for a unitary quantum channel using the minimal state set involving mixed, respectively pure, states. The generalization to noisy quantum channels is discussed in Sec.~\ref{sec:approaches_general}, where we also outline the reconstruction based on the Choi matrix, as a reference point for our later analysis. In Sec.~\ref{sec:application} we apply the reconstruction algorithms via numerical simulations to a physical model involving typical relaxation processes. We start with two-qubit examples and then investigate the extension to multiple qubits. In Sec.~\ref{sec:resource analysis} we compare the practical resource cost of the reconstruction algorithms in the context of state-of-the-art compressed sensing approaches, finding that it depends on whether the channel is close to unitary or experiences appreciable noise. Section~\ref{sec:conclusions} summarizes our findings.


\section{Reconstruction Approaches for Unitary Dynamical Maps}\label{sec:approaches_udm}

The action of a dynamical map on only two density matrices is sufficient to distinguish any two unitaries or to assess whether a given dynamical map is unitary~\cite{reich2013PRA}. Alternatively, it is possible, to employ a set of $d+1$ pure states, where $d$ is the dimension of the underlying Hilbert space~\cite{reich2013PRA}. This can be exploited to assess how well a quantum device realizes a desired prespecified operation with a minimal number of input states \cite{reich2013PRA}. We now show that the results of Ref.~\cite{reich2013PRA} can be extended to reconstruct an \textit{unknown} unitary operation $U$. Specifically, we demonstrate that the action of a dynamical map on at least two density matrices, respectively $d+1$ pure states, is sufficient to reconstruct the operation $U$ describing the action of a unitary dynamical map $\mathcal{D}_U(\rho) = U \rho U^\dagger$. We explain how $U$ can be obtained from the action of the dynamical map on the two minimal state sets mentioned above, making no further assumptions on the dynamical map except for its unitarity. This 
sets the stage for Sec.~\ref{sec:approaches_general} where we extend the reconstruction to non-unitary 
dynamical maps~\footnote{while we will make no specific assumptions on the dynamical map, the question of extracting its unitary part is meaningful only for dynamical maps sufficiently close to unitary.}.

\subsection{Reconstruction involving mixed input states}\label{ssec:approaches_udm_mixed} 

Reconstruction of $U$ for any unitary dynamical map, defined on a $d$-dimensional Hilbert space, is possible with only two input states. The first state is, in an arbitrary basis representation, 
\begin{equation}
\rho_B = \sum_{i=1}^d \lambda_i\ket{i}\bra{i}\,,\label{eq:rhoB mixed}
\end{equation}
where $\lambda_{i}\geq0$ and $\sum_{i}\lambda_{i}=1$ with $\forall i\neq j:\,\lambda_{i}\neq\lambda_{j}$. In words, we choose an arbitrary nondegenerate diagonal matrix with full rank and unit trace; it necessarily is a mixed state. The choice of $\rho_B$ allows the action of the unitary $U$ on the canonical basis to be reconstructed up to relative phases. The second state $\rho_{P}$ is given by 
\begin{equation}\label{eq:rhoP}
    \rho_P=\sum_{ij=1}^d \frac{1}{d} \ket{i}\bra{j}\,.
\end{equation}
Its matrix representation is constructed such that it is composed entirely of nonzero entries in the canonical basis which allows the missing relative phases to be obtained.

For the remainder of this work we choose the set of eigenvalues in Eq.~\eqref{eq:rhoB mixed} such that the distance of each $\lambda_{i}$ to the neighboring $\lambda_{i-1}$ and $\lambda_{i+1}$ is maximal and uniform for all $i$. Generally, for an unknown dynamical map, this ansatz is favorable since it protects well against the appearance of degeneracies in the output states due to noise. However, if only a single qubit decays in a multi-qubit system, then it might be more advisable to bias the choice towards the subspace of the decaying qubit.

Since unitary dynamical maps preserve the spectrum of density matrices, the image of $\rho_B$ can be written as
\begin{equation}
\mathcal{D}_U(\rho_B) = \sum_{i=1}^d \lambda_i \ket{\psi_i}\bra{\psi_i}\,,
\label{eq:Image mixed rhob}
\end{equation}
where the $\ket{\psi_i}$ form an orthonormal basis. The states $\ket{\psi_i}$ can be obtained
from diagonalization of $\mathcal{D}_U(\rho_B)$. Using these states, $U$ can always be written as (see Appendix~\ref{sec:appendix A} for details)
\begin{equation}
U = \sum_{k}e^{i\varphi_k}\ket{\psi_k}\bra{k}\,\label{eq:unitary_ansatz}
\end{equation}
with yet to be determined relative phases $\varphi_k$. Without loss of generality we can choose
$\varphi_1=0$, due to the insignificance of the global phase. We show in Appendix \ref{sec:appendix A} that the remaining phases can be determined from the image of $\rho_P$ and calculation of the
matrix elements
\begin{equation}
\varphi_{k}=\arg \left[d \bra{\psi_k}\mathcal{D}_{U}(\rho_{P})\ket{\psi_1}\right]\,,\label{eq:phase_reconstruction}
\end{equation}
for $k=2,\dots,d$.


\subsection{Reconstruction via pure input states}\label{ssec:approaches_udm_pure}

A caveat of the approach in Sec.~\ref{ssec:approaches_udm_mixed} is the difficulty to prepare mixed states. To amend this shortcoming we can replace $\rho_B$ by the set of pure states
\begin{equation}
    \rho^{(i)}_B=\ket{i}\bra{i}\,
    \label{eq:rhoB pure}
\end{equation}
for $i=1,\dots,d$, corresponding to the canonical basis. The state $\rho_P$ is pure and remains unchanged, cf. Eq.~\eqref{eq:rhoP}.

The reconstruction algorithm then proceeds as follows. Since unitary evolutions map pure states onto pure states, the image of $\rho^{(i)}_B$ can be written as
\begin{equation}
\mathcal{D}_U(\rho^{(i)}_B) = \ket{\psi_i}\bra{\psi_i}\,,\label{eq:Image pure rhob}
\end{equation}
where the $\{\ket{\psi_i}\}$ form an orthonormal basis. They can be obtained by either diagonalizing $\mathcal{D}_U(\rho^{(i)}_B)$ or applying $\mathcal{D}_U(\rho^{(i)}_B)$ to a random Hilbert space vector and renormalizing the resulting image. The latter only fails in rare cases where the random state is entirely in the kernel of $\mathcal{D}_U(\rho^{(i)}_B)$ which can be amended by choosing a different random state. Analogously to the reconstruction involving mixed input states, it can be shown that $U$ can then always be written as in Eq.~\eqref{eq:unitary_ansatz}
%
%
with yet to be determined relative phases $\varphi_k$.
From this point on the algorithm proceeds identically to the reconstruction
involving mixed input states since the state $\rho_P$ employed to obtain the phases is pure. In particular, the phases are obtained according to Eq.~\eqref{eq:phase_reconstruction}.


\section{Constructing Unitary Approximations for General Dynamical Maps}\label{sec:approaches_general}

In Sec.~\ref{sec:approaches_udm}, the reconstruction approaches assumed that the dynamical map is unitary. With some adjustments it is possible to obtain a unitary approximation to a general dynamical
map. We focus here on simplicity in application. Although the adjusted algorithms are not guaranteed to lead to the closest unitary to a given channel with respect to a certain distance measure, their numerical complexity is comparable to the approaches from Sec.~\ref{sec:approaches_udm} and their implementation is straightforward. Furthermore, we also introduce a reconstruction algorithm based on the Choi matrix.

\subsection{Construction involving mixed input states}\label{ssec:approaches_general_mixed}

We start again by considering the image of $\rho_B$ as defined in Eq.~\eqref{eq:rhoB mixed}. General dynamical maps may alter the spectrum of density matrices. Thus the eigenvalues $\lambda_i$ of $\rho_B$ are not necessarily eigenvalues of the image $\mathcal{D}(\rho_B)$. Provided the dynamical map is close to unitary, one can expect the
ordering of eigenvalues in $\mathcal{D}(\rho_B)$ to be preserved. This implies that by diagonalizing $\mathcal{D}(\rho_B)$ and sorting the eigenstates according to their eigenvalues, they can still be associated with the canonical basis states $\ket{i}$ needed in Eq.~\eqref{eq:unitary_ansatz}.  In the second step, $\rho_P$ is used as input state and Eq.~\eqref{eq:phase_reconstruction} provides an expression to construct the missing relative phases, in complete analogy to Sec.~\ref{ssec:approaches_udm_mixed}.
For strongly non-unitary dynamical maps, the approach might fail due to degeneracies in the output states.


\subsection{Construction via pure input states}\label{ssec:approaches_general_pure}

We start again by considering the image of the set $\rho^{(i)}_B$ as defined in Eq.~\eqref{eq:rhoB pure}. General dynamical maps do not necessarily map pure states to pure states, and a spectral decomposition of the image $\mathcal{D}(\rho^{(i)}_B)$ cannot be expected to yield one unique basis state $\ket{\psi_i}$. However, if the dynamical map is close to unitary, then the eigenstate with largest eigenvalue of $\mathcal{D}(\rho^{(i)}_B)$ should provide a reasonable ansatz for $\ket{\psi_i}$ in Eq.~\eqref{eq:unitary_ansatz}. As a second obstacle, it is not guaranteed  that the states $\ket{\psi_i}$ form an orthonormal basis which is a key assumption for the unitary reconstruction algorithm. It can be amended by performing a Gram-Schmidt orthonormalization on the set $\{\ket{\psi_i}\}$. The resulting orthonormal basis of the underlying Hilbert space can then be used in Eq.~\eqref{eq:unitary_ansatz} and the second step involving $\rho_P$ as an input state proceeds via Eq.~\eqref{eq:phase_reconstruction}, in complete analogy to Sec.~\ref{ssec:approaches_udm_pure}.


\subsection{Construction via the Choi matrix}\label{ssec:approaches_general_Choi}

The Choi matrix contains the complete information of a dynamical map. It is  therefore obvious that one can reconstruct the unitary from the Choi matrix. Since we have not found an explicit statement of the corresponding algorithm, we provide it here for completeness. 

In general, for Kraus operators $K_i$ such that 
\begin{equation}
\mathcal{D}(\rho)=\sum_i K_i \rho K^\dagger_i\,,
\label{eq:KraussDecomp}
\end{equation}
the Choi matrix $\chi_\mathcal{D}$ for the dynamical map $\mathcal{D}$ is given by
\begin{equation}
\chi_\mathcal{D}=\sum_{ij} |i\rangle \langle j| \otimes \mathcal{D}(|i\rangle \langle j|)\,,
\label{eq:DefChoi}
\end{equation}
where $\ket{i}$ are the canonical basis states. Then the Kraus operators can be obtained from diagonalizing $\chi_\mathcal{D}$~\cite{zyczkowski2004OSID}. For a unitary dynamical map, there only exists one Kraus operator, i.e., there is only one nonzero eigenvalue of the Choi matrix, and the corresponding eigenvector allows for reconstructing the unitary. An approach to obtain a unitary approximation to a general dynamical map via the Choi matrix is then to diagonalize the Choi matrix, select the eigenvector with largest modulus, and construct the corresponding Kraus operator $K_\text{max}$. This yields the closest unit-rank approximation to the Choi matrix. The resulting Kraus operator is, however, not necessarily unitary. This can be amended by a singular value decomposition of $K_\text{max}$ to obtain $K_\text{max}=VDW^\dagger$, and $U=VW^\dagger$ is the closest unitary in the operator norm to $K_\text{max}$. This provides a unitary approximation for the dynamical map via the Choi matrix. \correction{While the decomposition into unitary and non-unitary parts of a dynamical map is not unique, constructing the unitary $U=VW^\dagger$ from the dominant Kraus operator is a natural choice for weak dissipation.}


\section{Application}\label{sec:application}

Next we test and compare the reconstruction methods outlined in Sec.~\ref{sec:approaches_general}.  We begin by introducing the physical model in Sec.~\ref{subsec:Model} and introduce the reconstruction error $1-F_G$  in Sec.~\ref{subsec:fig of merit} as the main figure of merit to quantify the performance of the reconstruction methods. After analyzing the case of two qubits in Sec.~\ref{subsec:Two qubits}, we investigate  how the  performance of the methods scales with the number of qubits in Sec.~\ref{subsec:Multiple qubits}.

\subsection{Model}
\label{subsec:Model}
We illustrate the reconstruction algorithms for the scenario where the dissipative dynamics is described by a GKLS master equation~\cite{BreuerBook}, 
\begin{equation}
\dot{\rho}_S(t) = - i [H_0,\rho_S(t)] + \sum_i \gamma_i \Big( L_i \rho_S(t) L^\dagger_i - \frac{1}{2} \{L^\dagger_i L_i , \rho_S(t)\} \Big) \,,
\label{eq:Lindblad}
\end{equation}
with jump operators $L_i$, rates $\gamma_i$, and $H_0$ describing the coherent part of the evolution. 
The solution of the master equation at time $T_g$ provides a dynamical map, 
\begin{align}
    \rho(t=T_g) \equiv \mathcal{D}(\rho) \,,
    \label{eq:output dynmap}
\end{align}
where $T_g$ could be interpreted as, e.g., the gate time for realizing a particular unitary gate \correction{\footnote{We use the \textit{QuTiP} library to solve the GKLS master equation, setting absolute and relative error tolerances to $10^{-10}$.}}. We model the two most common relaxation processes, characterized by a relaxation time $T_1$, respectively dephasing time $T_2$, 
\begin{align}
\begin{split}
    &T_1: \hspace{0.5cm}L_1 = \sigma^-, 
    \hspace{0.5cm} \gamma_1 = \gamma_{\downarrow}=\frac{1}{T_1} \,,
\\
     &T_2: \hspace{0.5cm}L_2 = \sigma^z, \hspace{0.5cm} \gamma_2 = \gamma_{z}=\frac{1}{T_2} \,.  \end{split}
     \label{eq:T1undT2}
\end{align}
$T_1$ describes amplitude damping for a bath effectively at zero temperature, as it is typical in, e.g., superconducting qubit systems. $T_2$ characterizes pure dephasing in the $\sigma^z$ basis. Our choice of the rates is based on the fact that typical superconducting qubits are designed at frequencies of $\omega/2 \pi = 5 \text{GHz}$ and are operated at dilution refrigerator temperatures of $T \approx 20 \text{mK}$ \cite{Krantz2019}. In this limit, the up rate is negligible. For the Hamiltonian $H_0$, we consider two different scenarios: 
\begin{enumerate}[label=(\roman*)]
    \item We generate random unitaries $U_0 = U_\text{Rdm}$ on a collection of $N$ qubits, i.e. $d=2^N$, distributed according to the Haar measure \cite{mezzadri2007arXiv}, and obtain the Hamiltonian $H_\text{Rdm}$ 
    via the complex matrix logarithm,
    \begin{equation}
        H_\text{Rdm} =   \frac{i}{T_g}\log{ (U_\text{Rdm})}  = i \log{ (U_\text{Rdm})} \,,
        \label{eq:Def H_rdm}
    \end{equation}
 taking its main branch and setting the gate time $T_g$ to 1 for simplicity. 
\item  We consider the CNOT operation for superconducting (SC) qubits, 
 \begin{align}
     U_\text{SC}(t)= \mathcal{T} \exp {\Big(- i\int_0^t H_{SC}(t') dt' \Big)} \,,
     \label{eq:USC}
 \end{align}
generated by the cross-resonance Hamiltonian $H_\text{SC}(t)$ \cite{Devoret2010PhysRevB},
\begin{eqnarray}\nonumber
        H_\text{SC}(t) &=& -\frac{\omega_1}{2}\sigma^z_1 - \frac{\omega_2}{2}\sigma^z_2 + \frac{J}{2} \left( \sigma^x_1 \sigma^x_2 + \sigma^y_1 \sigma^y_2\right)\\&& + u_0(t) \cos{(\omega_2t)} \sigma^x_1 \,.
        \label{eq:H_SC}
 \end{eqnarray}
\correction{A single transversal drive on one of the two qubits is sufficient to  generate a CNOT gate~\cite{Zhang2003arXiv,Devoret2010PhysRevB}.} When accounting for dissipation, the time evolution in Eq.~\eqref{eq:USC} has to be replaced by the solution of Eq.~\eqref{eq:Lindblad} with the coherent part $H_0$ given by $H_\text{SC}$. 
 \end{enumerate}

\subsection{Figure of merit}
\label{subsec:fig of merit}

In order to assess the fidelity of the reconstruction, we employ the gate fidelity $F_G$ as our main figure of merit,
    \begin{equation}
        F_G(U_0,U_{RC})= \frac{1}{d} \left|\text{Tr}\{U_0^\dagger U_{RC} \}\right| \,,
        \label{eq:Gate_fid}
    \end{equation}
where $U_0$ denotes the unitary generated by the dynamics in the absence of any dissipation and $ U_{RC}$ denotes the reconstructed unitary. 

The fidelity between two arbitrary dynamical maps can be assessed at the level of the Choi matrix. Two particularly common choices are the process fidelity  and the average fidelity.
The process fidelity is given by~\cite{Magesan2012arXiv}
\begin{align}
    F_{\text{pro}}(\chi_0,\chi_{RC}) = \frac{1}{d^2} \text{Tr} \{ \chi_0 \chi_{RC} \} \,,
     \label{eq:Pro_fid}
\end{align}
where $\chi_0$ and $\chi_{RC}$ are the Choi matrices corresponding to the dynamical maps of the original unitary $U_0 \rho U_0^\dagger$ and the unitary approximation $U_{RC} \rho U_{RC}^\dagger$, respectively. 
The average fidelity $F_{\text{avg}}(\chi_0,\chi_{RC})$ is closely related to the process fidelity~\cite{horodecki1999arXiv},
    \begin{equation}
        F_{\text{avg}}(\chi_0,\chi_{RC}) = \frac{d F_{\text{pro}}(\chi_0,\chi_{RC})+1}{d+1} \,.
        \label{eq:Favg und Fpro}
    \end{equation}
Our results were found not to depend on the particular choice of gate, process, or average fidelity, and we only report the gate fidelity in Sec.~\ref{subsec:Two qubits} below.

Additionally, we characterize the non-unitarity of the dynamical map with
$1-u(\chi)$ where $u(\chi)$ is the so-called unitarity of the Choi matrix $\chi$ \cite{Wallman2015IOP,roth2018PRL},
    \begin{equation}
        u(\chi) = d^2 (d+1)^2 \hspace{0.1cm}\text{Var}[F_{\text{avg}}(\chi^k_{CL},\chi)] \,.
        \label{eq:Unitarity}
    \end{equation}
The variance in Eq.~\eqref{eq:Unitarity} is taken over unitary maps with corresponding Choi matrix $\chi^k_{CL}$ corresponding to a Clifford gate $U^k_{CL}$ randomly drawn from the Clifford group.  
We use 100 random Clifford gates to evaluate the variance of their average fidelites with respect to the Choi matrix. 

\subsection{Results: Two qubits}
\label{subsec:Two qubits}

\begin{figure}[tbp]
     \centering
               \includegraphics[width=0.45\textwidth]{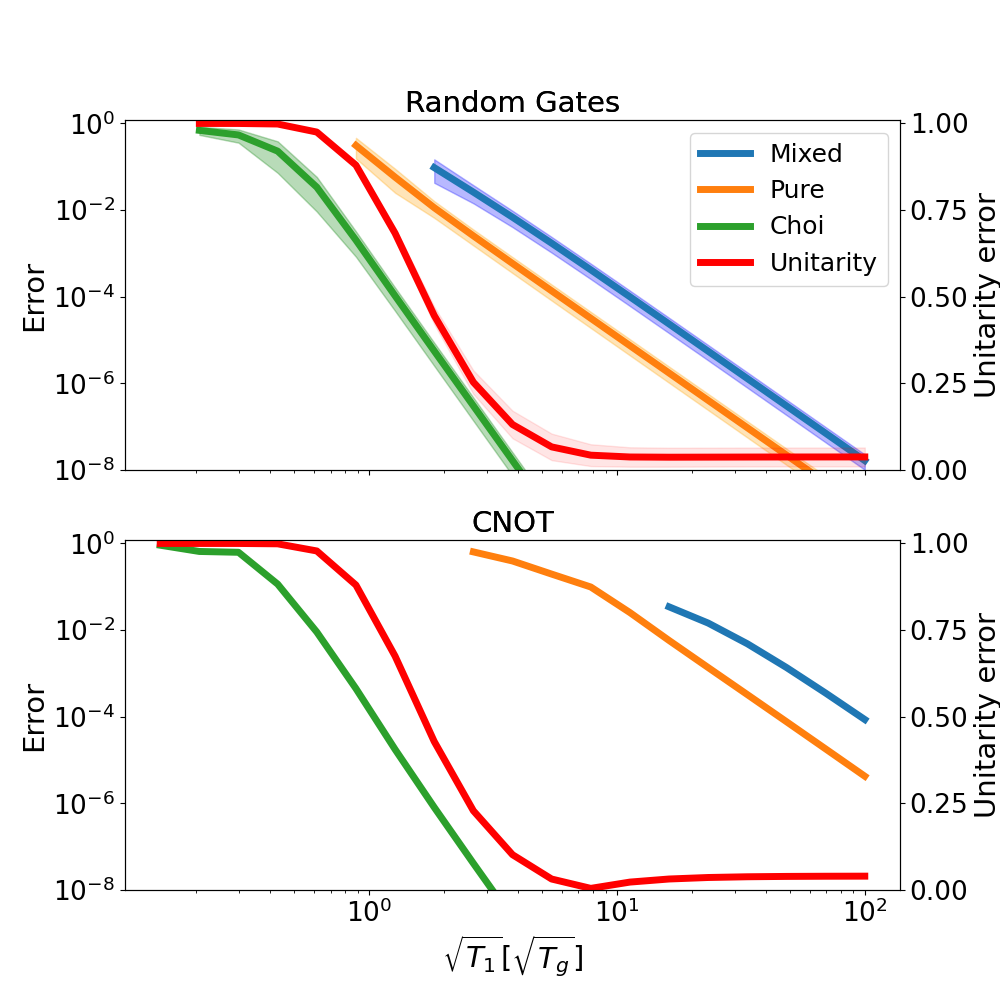}
           \caption{
           Reconstruction error $1-F_G(U_0,U_{RC})$ versus $T_1$ (left y-axis) of the reconstruction methods for random two qubit gates (left) and a CNOT gate (right). The shaded areas (left) represent the $1\sigma$ deviation with respect to the set of 100 random unitary gates drawn according to the Haar measure. The plot also shows the unitarity error $1-u(\chi)$ (right y-axis). Absorbing the decoherence rate $\gamma_k$ in Eq.~\eqref{eq:Lindblad} into the Lindblad operator leads to $L_{1}= \frac{1}{\sqrt{T_1}}\sigma^- \otimes \mathds{1}$ and $L_{2}= \frac{1}{\sqrt{T_1}} \mathds{1} \otimes \sigma^-$. The scaling $\sqrt{T_1}$ of each Lindblad operator is given in units of the gate time $\sqrt{T_g}$. }
         \label{fig:Fid over T1}
\end{figure}

Figure~\ref{fig:Fid over T1} shows the reconstruction error $1-F_G$ for the mixed-state, pure-state, and Choi reconstruction method versus the $T_1$ decay time for random two-qubit operations (top panel) as well as the CNOT gate (bottom panel). It also shows the unitarity error $1-u(\chi)$ of the Choi matrix $\chi$ corresponding to the full dynamical map $\mathcal{D}(\cdot)$ according to Eq.~\eqref{eq:output dynmap}. Note that the dissipation strength $\gamma_1$ of the dynamical map increases from right to left on the x-axis due to $\gamma_1 = \frac{1}{T_1}$. Figure~\ref{fig:Fid over T1} clearly demonstrates that the pure-state and mixed-state reconstruction methods can be employed for dynamics that are sufficiently close to unitary, cf. the unitarity error (red line). Even when the decay time $T_1$ decreases, i.e. the dissipation strength increases, and the unitarity error of the dynamical map becomes larger, the two methods are  able to approximately reconstruct the unitary part of the dynamical map for a wide range of dissipation strengths. 

The lines for the reconstruction error of the mixed-state and pure-state reconstruction do not cover the entire range of $T_1$ times. This is due to the breakdown of the methods for large deviations from unitarity: When the dissipation strength increases, the assumptions used in Sec.~\ref{sec:approaches_udm} are no longer fulfilled, and degeneracies in the spectrum of the images $\mathcal{D(\rho)}$ prohibit the use of the pure-state and mixed-state reconstruction methods. \correction{For a single CNOT gate in e.g. superconducting qubits, this is irrelevant: For example, on IBM Torino, with $T_1$ of the order of 100$\,\mu$s and $T_g\approx 100\,$ns~\cite{MiessenPRXQ24}, typical $T_1$ times are of the order of  $T_1 \sim 1000 T_g $ 
which results in reconstruction errors around $10^{-8}$ \footnote{Since $\sqrt{T_1} \sim 30 T_g$} and is far beyond the $T_1$ values where degeneracies occur. For several consecutive gates, however, one can define effective $T_1$ times which will be smaller such that degeneracies may occur eventually. } 
In contrast, the reconstruction via the Choi matrix is not influenced by degeneracies in the image. Hence the corresponding reconstruction errors stretch across the entire range of $T_1$ times. Figure~\ref{fig:Fid over T1} furthermore shows that, for the mixed-state reconstruction, degeneracies in the image of the dynamical map occur for smaller dissipation strengths compared to pure-state reconstruction (compare the $T_1$-coverage of the orange and blue lines in Fig.~\ref{fig:Fid over T1}). This is because the pure-state images in Eq.~\eqref{eq:Image pure rhob} are more protected against degeneracies than the mixed-state image (see Eq.~\eqref{eq:Image mixed rhob}). It reflects the fact that for pure initial states it is less likely that the dissipative part of the dynamical map annuls the spectral gap between  eigenvalues of different eigenvectors than it is for mixed initial states. 

For strong dissipation, the ordering of the eigenvalues and corresponding assignment to eigenstates may possibly get reshuffled. While there is not necessarily any degeneracy in the spectrum -- thus the reconstruction algorithm formally works -- the assignment will then not be faithful to the unitary to be reconstructed. In our simulations this regime is only reached beyond the point where degeneracies occurred such that the corresponding data could easily be discarded as not meaningful. In general, while this could be more difficult to identify, the unitarity of the dynamical map provides a way to estimate whether the key assumptions of the reconstruction method hold or not.

We observe in Fig.~\ref{fig:Fid over T1} that the mixed-state reconstruction performs worse in terms of the reconstruction error $1-F_G$ than the pure-state reconstruction, while the Choi reconstruction performs best. This behavior is intuitive if one considers the total amount of information available: The Choi matrix contains, via Eq.~\eqref{eq:DefChoi}, the information of the action of the dynamical map on arbitrary input states. The total amount of information accessible to the algorithm for the reconstruction is therefore higher compared to mixed-state and pure-state reconstruction. 
Similarly, since for mixed-state reconstruction a smaller number of states is propagated, less information is gained compared to the 
pure-state reconstruction. 
This explains the difference in reconstruction errors between mixed-state and pure-state reconstruction. 

\begin{figure}[tbp]
     \centering
         \includegraphics[width=0.45\textwidth]{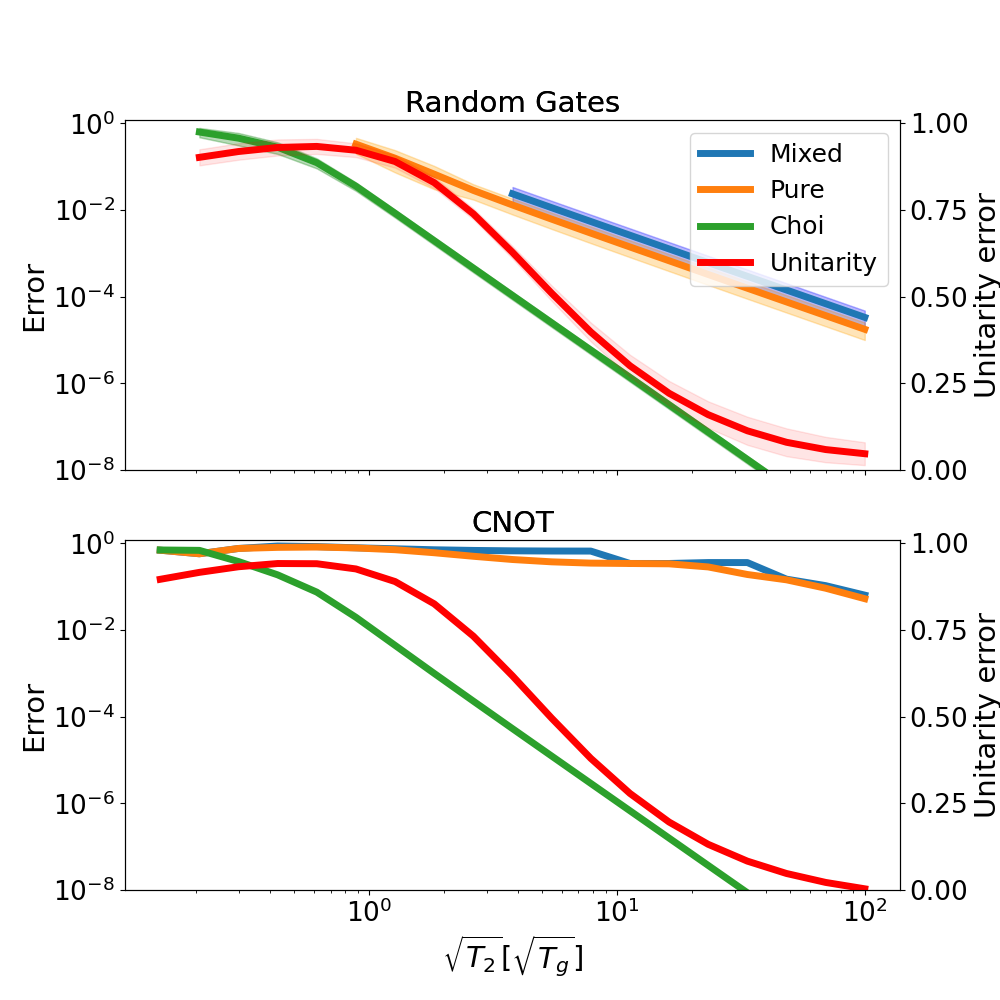}
           \caption{Same as Fig. \ref{fig:Fid over T1} but for $T_2$ dissipation instead of $T_1$.}
         \label{fig:Fid over T2}
\end{figure}

Figure~\ref{fig:Fid over T2} illustrates that the pure-state and mixed-state reconstruction methods can also be employed for pure dephasing, provided the dynamics is sufficiently close to being unitary. The image of the propagated pure initial states in the case of the CNOT gates is non-degenerate for the entire interval of $T_2$ times that we have considered, as seen by the continuous orange and blue lines in the lower panel of Fig.~\ref{fig:Fid over T2}.  For the CNOT there is no breakdown of the reconstruction algorithms for pure dephasing because the basis states in Eq.~\eqref{eq:rhoB mixed} and~\eqref{eq:rhoB pure} are not affected by the dephasing and the gate merely maps the basis states onto themselves. In contrast, a random unitary may rotate the basis states such that the output state likely are affected by the pure dephasing. 

For the mixed-state reconstruction of random gates, there is no appreciable difference in the general behavior as a function of $T_1$ and $T_2$, cf. the top panels of Fig.~\ref{fig:Fid over T1} and Fig.~\ref{fig:Fid over T2}. In particular, larger reconstruction errors compared to the pure-state reconstruction method are observed for both dissipation processes. 
For the CNOT gates, two observations can be made: The reconstruction errors for both mixed-and pure-state reconstruction are larger than for the random gates. The dissipation strength, at which the image becomes degenerate, is larger for pure dephasing compared to $T_1$ decay (cf. the ranges of the blue and orange lines in the lower panels of Fig.~\ref{fig:Fid over T1} and Fig.~\ref{fig:Fid over T2}). This is in contrast to the random gates, where there is no significant difference between $T_1$ and $T_2$ for both mixed-state and pure-state approach.

\subsection{Results: Multiple qubits}
\label{subsec:Multiple qubits}
\begin{figure}[tbp]
\centering\includegraphics[width=0.49\textwidth]{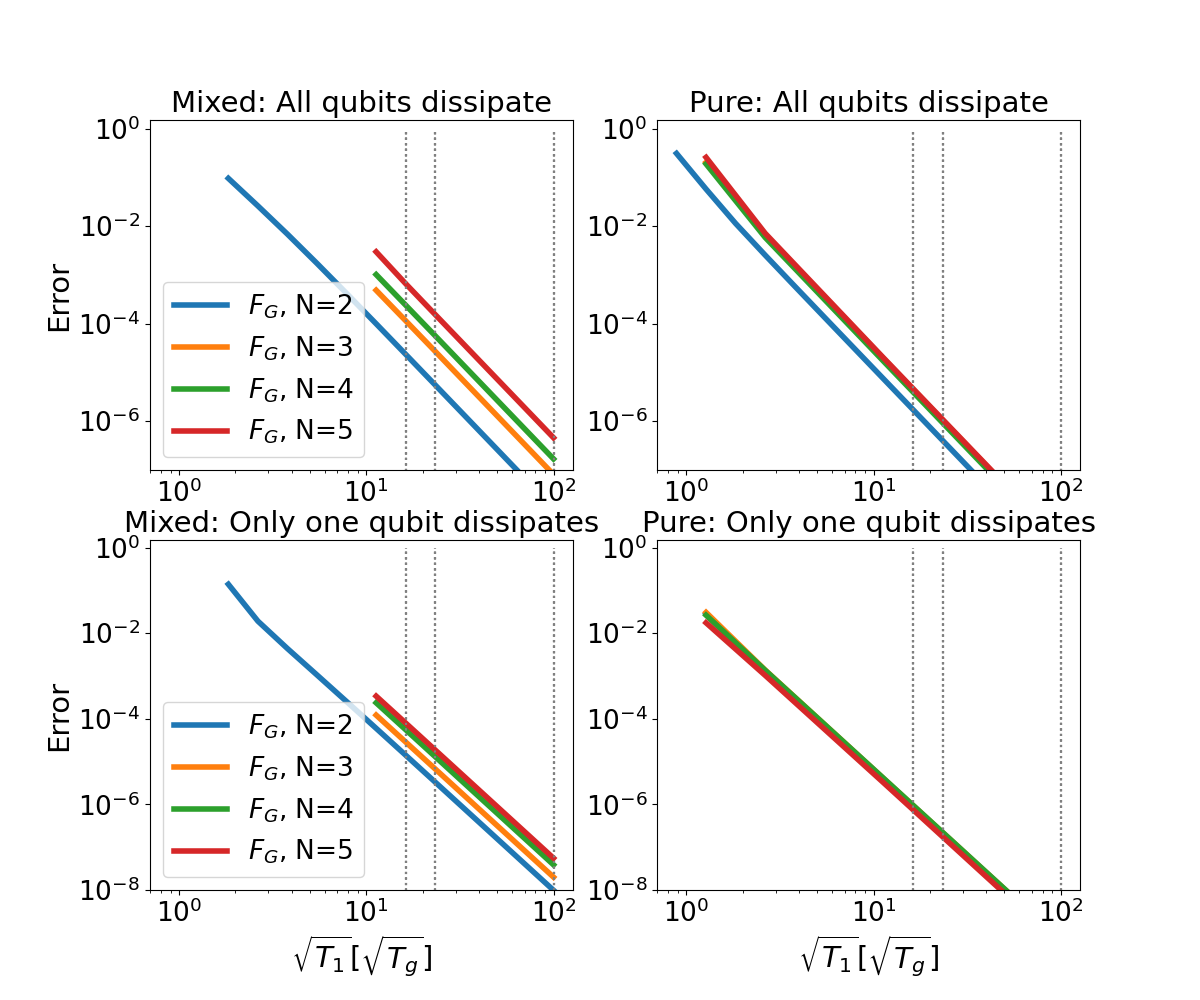}
           \caption{Reconstruction error $1-F_G(U_0,U_{RC})$
           averaged over $n=\{100,75,50,10\}$
           random unitary operations of the mixed-state reconstruction method versus $T_1$ decay time for $N=\{2,3,4,5\}$ qubits (for example for three qubits, i.e, $N=3$, the average reconstruction error of $n=75$ different random unitary operations is calculated). The dissipation on the additional qubits with respect to a single qubit is (left) identical to the coupling of the first qubit, respectively (right) set to zero, i.e., only one qubit exhibits dissipation.}
         \label{fig:error over RDM T1 for N}
\end{figure}

How does the system size affect the performance of the reconstruction methods? To answer this question, we consider random unitaries on a system of $N$ qubits for the case of $T_1$ decay. \correction{Since unitaries defined on the full Hilbert space are considered, almost all of them are entangling, i.e. their generation requires a physical Hamiltonian with at least two-body interactions.} We have seen in Sec.~\ref{subsec:Two qubits} that the reconstruction error of the reconstruction methods is connected to the unitarity of the dynamical map. The specific form of the dissipation determines unitarity, and we consider two choices in the following. In the first choice, all qubits are subject to decay, with identical dissipation rates $\gamma_1^{(i)} = \frac{1}{T_1}$ for $i=1,\dots,N$. Such a scenario is common in qubit arrays where the error is caused by spontaneous decay~\cite{Schindler2013arXiv}. In this case, one would expect the reconstruction error to increase with system size, since the total dissipation strength is $\gamma_1^{\text{tot}}= \sum_i^N \gamma_1^{(i)}$ which scales with $N$. Figure~\ref{fig:error over RDM T1 for N} (top) shows the reconstruction error versus $T_1$ times for random unitaries for $N=2,3,4,5$ qubits, reconstructed using the mixed-state (left panels), respectively the pure-state (right panels), reconstruction method. The top panels confirm the intuition that the reconstruction error gradually increases with the system size for all dissipation strengths. This gradual shift is more pronounced for mixed-state reconstruction (top left panel of Fig.~\ref{fig:error over RDM T1 for N}) than for the pure-state reconstruction (top right panel of Fig.~\ref{fig:error over RDM T1 for N}). 

In the second scenario,  only a single qubit is subject to dissipation, with rate $\gamma_1 = \frac{1}{T_1}$ (bottom panels in Fig.~\ref{fig:error over RDM T1 for N}). The dissipative qubit could be, e.g., a working qubit that is connected to memory qubits which do not exhibit the same level of noise \cite{Schindler2013arXiv}. Figure~\ref{fig:error over RDM T1 for N} (bottom) shows once again a gradual shift of the reconstruction error with system size for the mixed-state reconstruction. This shift is smaller if only one qubits dissipates compared to the case where all qubits dissipate, cf. the top and bottom left panels of Fig.~\ref{fig:error over RDM T1 for N}. The reconstruction error of the pure-state reconstruction does not change appreciably  if only one qubit dissipates, in contrast to the case where all qubits dissipate, cf. the top and bottom right panels of  Fig.~\ref{fig:error over RDM T1 for N}. Specifically, the reconstruction error increases with $N$ visibly when all qubits are subject to dissipation but remains almost constant for single qubit decay.

\begin{figure}[tbp]
     \centering
         \includegraphics[width=0.45\textwidth]{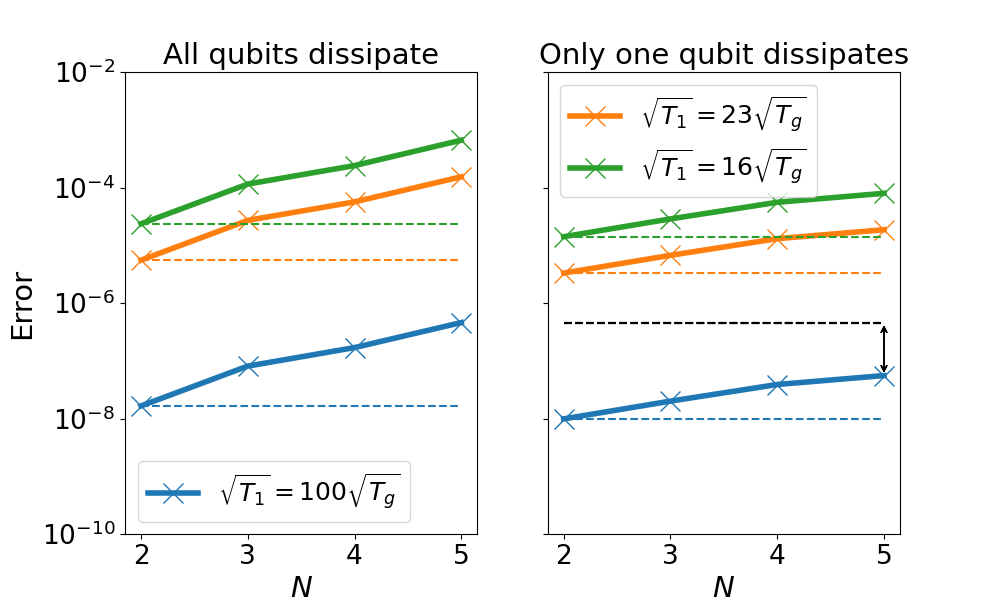}
           \caption{Behavior of reconstruction error $1-F_G(U_0,U_{RC})$ of the mixed-state reconstruction method for the three $T_1$ times indicated by grey vertical dashed lines in Fig.~\ref{fig:error over RDM T1 for N}.}
         \label{fig:mixed error over N for T1}
\end{figure}

\begin{figure}[tbp]
     \centering
    \includegraphics[width=0.45\textwidth]{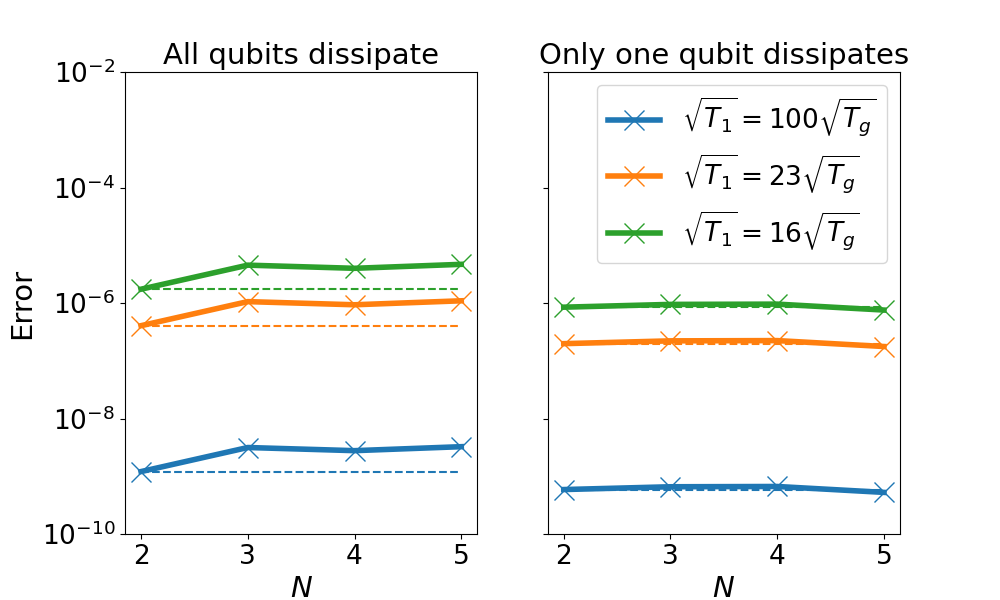}
           \caption{Same as Fig.~\ref{fig:mixed error over N for T1} but for pure-state reconstruction instead of the mixed-state approach. }
         \label{fig:pure error over N for T1}
\end{figure}

Since the differences of the reconstruction error are relatively small on the scale used in Fig.~\ref{fig:error over RDM T1 for N}, we analyze this behavior in greater detail for selected fixed $T_1$ decay times, see Fig.~\ref{fig:mixed error over N for T1} for mixed-state and Fig.~\ref{fig:pure error over N for T1} for pure-state reconstruction.   
 Figure~\ref{fig:mixed error over N for T1} confirms our observation from Fig.~\ref{fig:error over RDM T1 for N} that for the mixed-state reconstruction, the reconstruction error increases as a function of system size $N$ for both choices of dissipation and all $T_1$ decay times considered. The stronger scaling of the reconstruction error with the system size in the mixed-state reconstruction, which persists even if only one qubit is affected by the decay, can be explained by the increased likelihood of a degeneracy to occur in the image of the mixed input state $\rho_B$: When the dimension of the Hilbert space increases, the eigenvalues of $\rho_B$ move closer together since $\text{Tr}[\rho_B]=1$. Moreover, even if no degeneracies occur, the ordering of the eigenvalues of the output states is more susceptible to errors for larger system sizes. 
The increase of the error with $N$ in 
Fig.~\ref{fig:mixed error over N for T1} is much smaller when only one qubits dissipates, compared to the case where all qubits dissipate. This is indicated by the black dashed horizontal line in the right panel of Fig.~\ref{fig:mixed error over N for T1}. 

For the pure-state reconstruction, the left panel Fig.~\ref{fig:pure error over N for T1} shows that if all qubits dissipate, the reconstruction error increases with system size $N$. Conversely, if only one qubit dissipates, cf.~the right panel in Fig.~\ref{fig:pure error over N for T1}, the reconstruction error of the pure-state reconstruction is roughly independent of the system size, with small fluctuations explained by the randomness in drawing the unitarities to be reconstructed in our simulations.
This highlights that the consistently stronger performance of the pure-state reconstruction compared to the mixed-state reconstruction, observed in Sec.~\ref{subsec:Two qubits}, persists and becomes more pronounced for larger system sizes.

We have also analyzed the behavior of our reconstruction methods for CNOT-like gates for multiple qubits, see Appendix \ref{sec:appendix B}. Once again, we find the pure-state reconstruction to perform better than the mixed-state reconstruction in terms of both accuracy and range of dissipation strength where the method can be meaningfully used.

\subsection{Results: Influence of SPAM \correction{and coherent errors}}
\label{subsec:spam error}

So far, we have assumed that all input states are prepared perfectly and that no read-out errors occur. We now investigate the influence of state preparation and measurement errors (SPAM) on the performance of the reconstruction methods. To this end, we adopt the model describing SPAM errors as unwanted bit-flip operations on both input- and output states~\cite{Finsterhoelzl2023QST, Funcke2022PRA}. This yields the map 
\begin{equation}
\Phi[\rho] = (1-p) \rho + p \sigma^x \rho \sigma^x\,,
\label{eq:CPTP_spam}
\end{equation}
where $p$ denotes the probability that a SPAM error occurs. We apply $\Phi$ to the input states in Eqs.~\eqref{eq:rhoB mixed}, \eqref{eq:rhoP} and \eqref{eq:rhoB pure} to account for state preparation errors. Similarly, to model measurement errors, we apply the same map to all output states of the dynamical map (e.g. $\mathcal{D}(\rho_B)$ for the mixed initial state $\rho_B$).

\begin{figure}[tbp]
     \centering
    \includegraphics[width=0.45\textwidth]{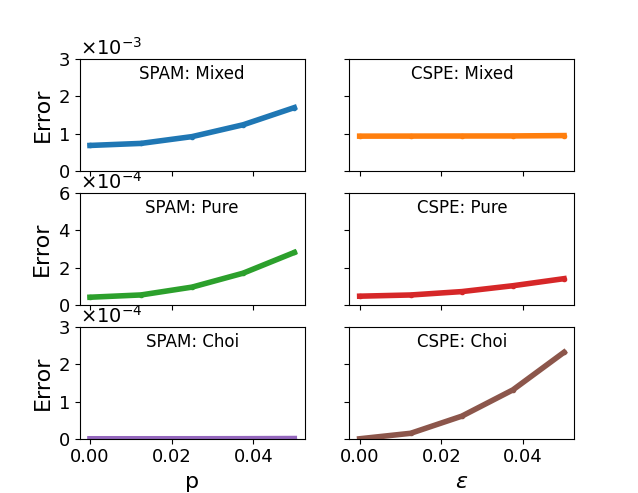}
           \caption{\correction{Reconstruction error 
           as a function of the SPAM error probability $p$ (left panel), cf. Eq.~\eqref{eq:CPTP_spam}, 
           resp. the strength $\epsilon$ of coherent state preparation errors (CSPE, right panel), cf.~Eq.~\eqref{eq:U_coh_err}, for amplitude damping with $\sqrt{T_1}=0.7\sqrt{T_g}$.
The results are averaged over 100 random unitary two-qubit gates for SPAM errors, resp.~over 100 random coherent rotations for CSPE.}}
         \label{fig:spam_error}
\end{figure}
\correction{The left panel in} Figure~\ref{fig:spam_error} shows how the reconstruction error changes for increasing SPAM error probability $p$, assuming a moderate dissipation strength of $\sqrt{T_1}=0.7\sqrt{T_g}$. For $p=0$, the reconstruction error in Fig. \ref{fig:spam_error} is the same as in Fig.~\ref{fig:Fid over T1} at the respective $T_1$ value. 
The moderate increase of the reconstruction error by at most one order of magnitude at the comparatively high SPAM error probability of 5\% indicates an overall robustness of all three methods to SPAM errors. Perhaps unsurprisingly, pure-state and Choi reconstruction are somewhat more sensitive to SPAM errors than the mixed-state reconstruction. Importantly, for SPAM errors up to 1\%, errors in the estimated output states do not seem to propagate to errors in the reconstructed unitary approximation, as the increase in the reconstruction error in this range is essentially negligible for all three methods in Fig.~\ref{fig:spam_error}.

\correction{Another error source in real setups are coherent state preparation errors (CSPE). To estimate their influence, we consider a Hamiltonian $H= \sum_{ij=0}^3 c_{ij} \sigma^i \otimes \sigma^j$ where each coefficient is drawn randomly, $0 \leq c_{ij} \leq 1$. Normalizing $H$ such that $\epsilon$ dictates the scale of the perturbation, this generates a random coherent rotation,
\begin{align}
    U = e^{-i \epsilon H}, \hspace{1cm} \epsilon \ll 1
    \label{eq:U_coh_err}
\end{align}
which is applied to the input states in Eqs.~\eqref{eq:rhoB mixed}, \eqref{eq:rhoP} and \eqref{eq:rhoB pure} to mimick CSPE. The right panel in Fig.~\ref{fig:spam_error} shows the change of the reconstruction error for the CNOT gate in Eq.~\eqref{eq:USC} at $t=T_g$ for increasing perturbation strengths $\epsilon$, assuming moderate dissipation, $\sqrt{T_1} = 7 \sqrt{T_g}$. Note that for $\epsilon=0$, the reconstruction error is identical to the one in Fig.~\ref{fig:Fid over T1} at the respective $T_1$ value.  While the error of the mixed-state reconstruction appears to be insensitive to coherent noise, an increase in the error (albeit on different scales) is observed for pure-state and Choi reconstructions. The different behavior of pure states compared to mixed states can be rationalized in the Bloch sphere picture: While the rotation angle is identical for both kinds of states, the absolute distance covered by the rotation is smaller for mixed states due to their smaller radii. Regarding the input "states" $\tilde{\rho}$ for the Choi reconstruction, Eq.~\eqref{eq:DefChoi}, a significant part of them is traceless, i.e., they only consist of coherences which may explain their larger sensitivity to coherent errors than to dissipative spam errors. Overall, Fig.~\ref{fig:spam_error} suggests all three methods to be rather robust with respect to both SPAM and coherent state preparation errors.}

\begin{figure}[tbp]
     \centering
    \includegraphics[width=0.45\textwidth]{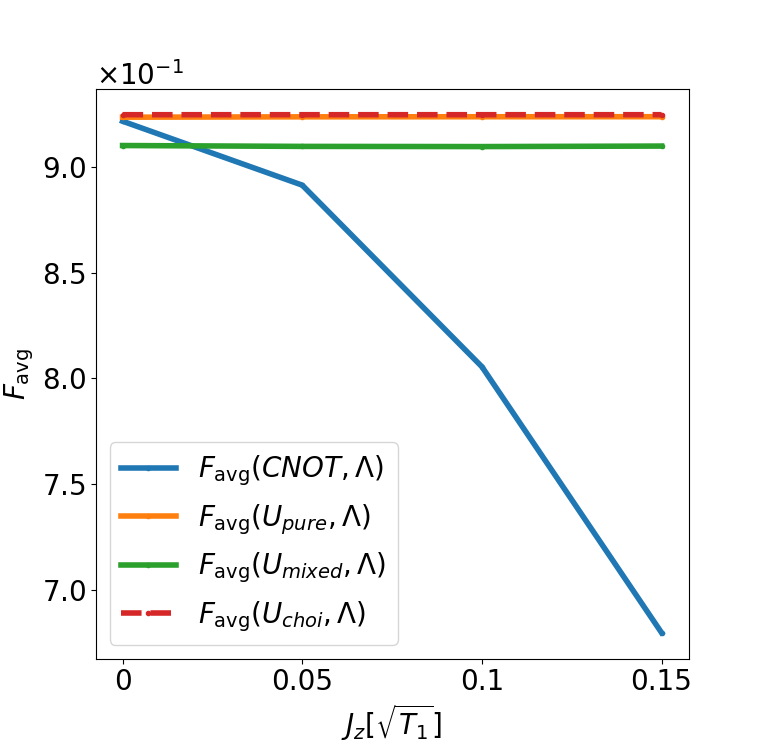}
           \caption{\correction{Accounting for crosstalk: Average gate fidelity, cf.~Eq.~\eqref{eq:Favg und Fpro}, as a function of the crosstalk strength $J_Z$, cf. Eq.~\eqref{eq:CPTP_spam}, 
           for a CNOT gate and amplitude damping with $\sqrt{T_1} = 26 \sqrt{T_g}$. }}
         \label{fig:coh_error}
\end{figure}
\correction{Lastly, we consider the influence of another kind of a coherent error, namely crosstalk. To this end, we modify the Hamiltonian~\eqref{eq:H_SC} generating the CNOT gate by adding a crosstalk term,}
\begin{equation}
    H_\text{CT}= J_z \sigma^z_1 \sigma^z_2, \hspace{1cm} H_\text{tot} = H_\text{CNOT} + H_\text{CT}\,.
\end{equation}
\correction{
As a result, for $J_z \neq 0$, a different gate is obtained at $t=T_g$, i.e., the unitary part of the CPTP map $\Lambda$ that translates the input into output states is not a CNOT. This is directly reflected in the blue curve in Fig.~\ref{fig:coh_error}, showing the average gate fidelity, cf.~Eq.~\eqref{eq:Favg und Fpro}, for CNOT. 
Importantly, Fig.~\ref{fig:coh_error} also shows that the average gate fidelity, when evaluated for the unitary part of the CPTP map instead of CNOT, is insensitive to the crosstalk strength. This is expected because crosstalk is a coherent error that can be absorbed into the unitary part of the evolution. It suggests to use the reconstruction method to determine the actually implemented gate and then account for the modification of the gate directly in the compilation of quantum circuits, rather than correct for the crosstalk.}

\section{Resource analysis}\label{sec:resource analysis}

Finally we quantify the resources needed for the reconstruction schemes and compare with alternative approaches. 
In order to reconstruct the unitary part $\mathcal{D}_U(\cdot)$ of the unknown dynamical map $\mathcal{D}(\cdot)$ using the reconstruction schemes, it is necessary to obtain the images of the different initial states under the dynamical map. In an experiment, these images need to be determined using quantum state tomography. Standard approaches for state tomography require $m_{\psi}=d^2$ measurement settings for a $d\times d$ density matrix (with $d=2^N$ for an $N$-qubit system)~\cite{Mohseni2008arXiv}. Modern state tomography methods such as compressed sensing~\cite{Gross2010arXiv,Flammia_2012} allow to recover a density matrix of rank $r \leq d$ from a smaller number of measurements $m = \mathcal{O} (rd)$ of randomly chosen Pauli expectations~\cite{Gross2010arXiv}. In the limit of a full rank density matrix $r=d$, this reduces to the number of measurement settings for standard state tomography. 

In practice, measurements for each setting have to be repeated a certain number of times to obtain sufficient measurement statistics. Our state-based reconstruction schemes involve a different number of states which has to be taken into account when analyzing their performance in terms of resources. Therefore, the suitable figure of merit to quantify the total resource scaling is the minimum number of channel uses $M$. To obtain $M$, we need to sum over the number of measurements for each state in the scheme. We now explain how we calculate the total resource scaling in greater detail.

The so-called sampling complexity $t$ accounts for the total number of measurements 
to obtain an estimate for a certain state, respectively channel. For example, the sampling complexity for state tomography of a full-rank density matrix is $t = \mathcal{O}(d^4) =\mathcal{O}(16^N) $ \cite{Gross2010arXiv}. Compressed sensing allows one to reduce the sampling complexity $t$ of a density matrix with rank $r$ to \cite{Flammia_2012}
\begin{align}
    t_{CS}(r) = \mathcal{O}(r^2 d^2) =\mathcal{O}(r ^2 4^N)\,.
    \label{eq:sampling comp cs}
\end{align}
Since it is the rank of the output density matrix that matters for the reconstruction,  we set $r \equiv r_{\text{out}}$. In general, the rank of each output state obtained from the non-unitary time evolution of the mixed, respectively pure initial states may be different. \correction{Since the sample complexity depends on the rank of the output states, a worst case analysis considers the maximum rank among all output states}, 
\begin{align}
    r_{\text{out}}=\text{max}\{\text{rank}(\mathcal{D}(\rho)) \}\,.
\end{align}
\correction{For the pure state reconstruction, the worst-case scenario corresponds to at least one rank 1 input state evolving to a full rank output state such that $r_\text{out}=d$. 
For the mixed state reconstruction, the output state automatically also has full rank and is therefore covered by the worst-case bound \footnote{If the output state is not of full rank then the mixed reconstruction cannot be used. This is however only possible if the dissipative part effectively purifies the input state such that $r_\text{out} < r_\text{in}=d$.}.}
Equation~\eqref{eq:sampling comp cs} captures the sampling complexity of a single output state which, for the overall resource estimate of the reconstruction, needs to be multiplied by the number of output states, $n_{\text{out}}$, required by the respective reconstruction method.
In this respect, it is possible to ignore the contribution of the state $\rho_P$ to the resource scaling for several reasons: (i) $\rho_P$ is needed for both pure-state and mixed-state reconstruction, so when comparing the resource scaling of both, its contribution can be discarded. (ii) Since $\rho_P$ is a single pure state, independent of the system size $N$, its contribution to the overall resource scaling is comparatively small.
(iii) There are no significant state preparation costs for $\rho_P$ as it involves only single qubit Hadamard gates on each qubit which in most modern architectures.
Neglecting the contribution of $\rho_P$, 
$n^{\text{Mix}}_{\text{out}}=1$ for the mixed-state reconstruction, corresponding to the  output state $\mathcal{D}(\rho_B)$. In contrast, the pure-state reconstruction has $n^{\text{Pure}}_{\text{out}}=d$ output states, namely $\mathcal{D}(\rho_B^{(i)})$ with $i=1,\ldots,d$.

Given $n_\text{out}$ and an estimate of the output states' rank, the minimum number $M$ of channel uses is
\begin{align}
    M(r_{\text{out}}) = \mathcal{O}(n_{\text{out}} \cdot t_{CS}(r_{\text{out}}))\,.
    \label{eq:channel uses mixed and pure}
\end{align}
We now evaluate this more concretely for mixed-state and pure-state reconstruction.
While the rank of the mixed input state is $r_{\text{in}}=\text{rank} (\rho_B)=d$, 
$r_{\text{out}} =\text{rank} (\mathcal{D}(\rho_B)) \leq d$ where the inequality occurs for strongly non-unitary dynamical maps. Since, however, the mixed-state reconstruction breaks down if the output states are not full rank, as shown in Fig.~\ref{fig:Fid over T1}, 
we only consider $r_{\text{out}} = d$ in the following. 
With Eqs.~\eqref{eq:sampling comp cs} and~\eqref{eq:channel uses mixed and pure}, the minimal number of channel uses for the mixed-state reconstruction then becomes
\begin{equation}
     M_{\text{Mix}}^{(U)}= \mathcal{O}(d^4)=\mathcal{O}(16^N) = M_{\text{Mix}}\,.
   \label{eq:channel uses mix unit}
\end{equation}
Here, the superscript $(U)$ in $M_{\text{Mix}}^{(U)}$ symbolizes that the dynamics are unitary or sufficiently close to unitary such that the rank of the output states is not appreciably affected. The absence of this superscript indicates the general case of (potentially strong) non-unitary dynamics. 
We assume that the non-unitary dynamical map is faithful in the sense that the full rank of the mixed input density matrix is preserved and no degeneracies occur. In this case, the number of channel uses for the mixed-state reconstruction does not change for unitary or non-unitary dynamics, i.e. $M_{\text{Mix}}^{(U)} = M_{\text{Mix}}$.

We now turn to the resource scaling of the pure-state reconstruction. The pure input states have rank $r^{(i)}_{\text{in}}=\text{rank} (\rho_B^{(i)})=1$. If the dynamical map is perfectly unitary $\mathcal{D}(\cdot) \equiv \mathcal{D}_U(\cdot)$ or sufficiently close to unitary $\mathcal{D}(\cdot) \approx \mathcal{D}_U(\cdot)$, then $r^{(i)}_{\text{out}} = \text{rank} (\mathcal{D}_U(\rho_B^{(i)}))=1$. We therefore obtain
\begin{equation}
 M_{\text{Pure}}^{(U)}= \mathcal{O}(d^3)=\mathcal{O}(8^N)\,.
    \label{eq:channel uses pure unit}
\end{equation}
If the dynamical map is non-unitary, $r^{(i)}_{\text{out}} =\text{rank} (\mathcal{D}(\rho_B^{(i)})) \geq 1$ such that
\begin{equation}
   M_{\text{Pure}}= \mathcal{O}(r_{\text{out}}^2 d^3)=\mathcal{O}(r_{\text{out}}^2 8^N)\,.
    \label{eq:channel uses pure}
\end{equation}
For the Choi reconstruction in Eq.~\eqref{eq:DefChoi}, the minimum number of channel uses  is $M_\text{Choi}= \mathcal{O}(N \hspace{0.1cm} 16^N)$ \cite{Mohseni2008arXiv}. The most efficient way to implement the Choi reconstruction uses compressed sensing \cite{Gross2010arXiv} and direct fidelity estimation \cite{Flammia2011arXiv}. 
Under the assumption of a rank-1 Choi matrix, the number of channel uses then reduces to~\cite{roth2018PRL}
\begin{equation}
    M^{(U)}_\text{Choi} = \mathcal{O}(d^4) = \mathcal{O}(16^N)\,.
    \label{eq:channel uses choi}
\end{equation}

Now we are able to compare the number of channel uses for the three methods. We consider two different cases for the output states: (i) The dissipative part of the dynamics is weak in the sense that one can approximate $r_{\text{out}} \approx r_{\text{in}}$. A comparison of Eqs.~\eqref{eq:channel uses mix unit}--\eqref{eq:channel uses choi} yields
\begin{align}
    M_{\text{Pure}}^{(U)} <  M_{\text{Mix}}^{(U)} \approx M^{U}_\text{Choi}\,.
    \label{eq:Vergleich_Channeluses_unit}
\end{align}
We conclude that the pure-state reconstruction requires the least amount of channel uses when the dynamics is unitary or close to unitary and is therefore the best choice in terms of required resources. 
(ii) The dissipative part strongly affects the rank of the output states such that every output state has approximately full rank, $r_{\text{out}} = d$. In this case, Eq.~\eqref{eq:channel uses choi} cannot be used since the rank-1 assumption is violated, and an estimate of the scaling for the Choi reconstruction 
is difficult~\cite{roth2018PRL,Flammia_2012}. We will thus not compare with compressed sensing estimates of $M_{\text{Choi}}$ for this scenario, and a comparison of Eq.~\eqref{eq:channel uses mix unit} with Eq.~\eqref{eq:channel uses pure} shows
\begin{align}
   M_{\text{Mix}} < M_{\text{Pure}}\,.
\end{align}
For strong dissipation, the mixed-state reconstruction outperforms the pure-state reconstruction in terms of channel uses. Intuitively, this follows from the fact that with stronger dissipation, the number of channel uses for the pure-state construction increases whereas it remains constant for the mixed-state reconstruction. Despite its worse scaling, there are three  reasons why the pure-state reconstruction may still be the better choice also for strong dissipation: First, the reconstruction error of the mixed-state reconstruction is higher than the pure-state one. Second, mixed states are experimentally much harder to initialize than pure states. Finally, the mixed-state reconstruction is more susceptible to degeneracies in the image, limiting the regime for which it can be used in practice. 

Finally, comparing Eq.~\eqref{eq:channel uses choi}, representing the state of the art in channel reconstruction, with the scaling of the pure-state reconstruction in Eq.~\eqref{eq:channel uses pure}, we assert that our method performs significantly better if $r_{\text{out}} \approx 1$ since the minimal number of channel uses becomes $M_{\text{Pure}}= \mathcal{O}(8^N) \ll \mathcal{O}(16^N) $. Even for more substantial dissipation, when  $r_{\text{out}} \approx \sqrt{d}$, both methods perform equally well since $M_{\text{Pure}}= \mathcal{O}(2^N8^N) = \mathcal{O}(16^N) = M^{(U)}_\text{Choi}$. Hence, our method decreases the number of channel uses necessary to reconstruct the unitary part of a general dynamical map compared to previous work using compressed sensing, as long as $r_{\text{out}} \leq \sqrt{d}$, at the cost of larger reconstruction errors compared to Choi reconstruction. When comparing with Standard Quantum Process Tomography (SQPT), which requires $M_\text{SQPT}= \mathcal{O}(N 16^N)$ channel uses \cite{Mohseni2008arXiv}, the reduction in the required resources is even more substantial since $M^{(U)}_\text{Choi} < M_\text{SQPT}$.

\begin{figure}[tbp]
     \centering
    \includegraphics[width=0.45\textwidth]{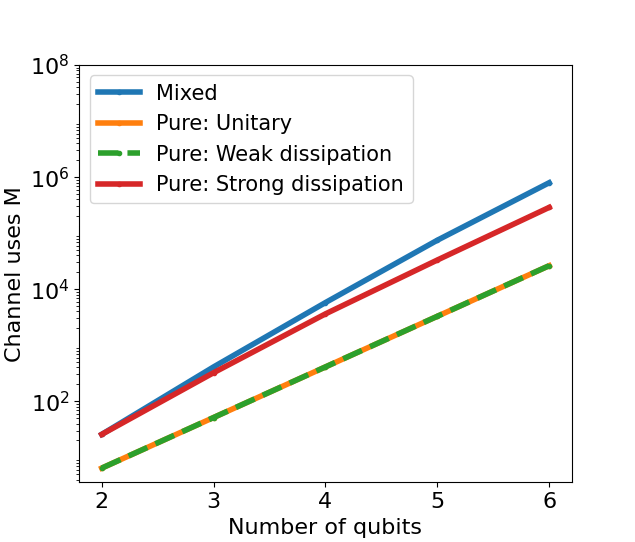}
           \caption{Minimum number of channel uses $M$ required for the pure-state and mixed-state reconstruction versus the number of qubits for different dissipation strengths. To avoid any bias due to the rank of the output states, $M$ is averaged over $n=\{100,75,50,10,10\}$ random unitaries for $N = \{2,3,4,5,6\}$ qubit assuming identical dissipation for each gate.  Weak dissipation corresponds to $\sqrt{T_2}= 25 \sqrt{T_g}$ (green) and strong dissipation to $\sqrt{T_2} = 7 \sqrt{T_g}$ (red); the number of channel uses for the mixed-state reconstruction is not affected by the dissipation strength. 
           The threshold  below which SVD values of the output states are considered zero was taken to be $\epsilon = 0.01$.} 
         \label{fig:MversusN}
\end{figure}
The advantage of the pure-state and mixed-state reconstruction methods compared to the Choi reconstruction results from exploiting the near-unitarity of the dynamical map. It is an advantage not only in terms of channel uses, as expressed in Eq.~\eqref{eq:Vergleich_Channeluses_unit}, but also in terms of ease of implementation, as our reconstruction methods do not require the preparation of entangled input states. They share this feature with low-rank process tomography methods~\cite{Kliesch2019quantum} which suggests to compare to these protocols in terms of both implementation and scaling. A key difference of our approach is that no optimization, typically carried out via a semi-definite program~\cite{Kliesch2019quantum}, is required; rather Eqs.~\eqref{eq:unitary_ansatz} and~\eqref{eq:phase_reconstruction} provide explicit formulae for reconstructing the unitary directly from the output states.

In order to compare the scaling of the number $M$ of channel uses as a function of system size, we proceed in two steps. First, 
we determine $M$ for each method with Eq.~\eqref{eq:channel uses mixed and pure} for different pure dephasing strengths. We only consider pure dephasing here as the worse case since amplitude damping may  
decrease the rank of output states $r_\text{out}$ and thus result in better scaling.
Since $r_\text{out}$ is the determining factor for the minimum number of required channel uses $M$ in Eq.~\eqref{eq:channel uses mixed and pure}, we extract it numerically 
for noisy randomly chosen $N$-qubit operations.
The resulting numbers of channel uses are shown in Fig.~\ref{fig:MversusN}, confirming the scaling predictions of Eqs.~\eqref{eq:channel uses pure unit} and \eqref{eq:channel uses pure}:
For dynamics close to unitarity, i.e.~weak dissipation, the rank of the output states does not change substantially such that the number of channel uses of the pure state reconstruction is identical to the purely unitary case. Furthermore, Fig.~\ref{fig:MversusN} also demonstrates that the prediction of Eq.~\eqref{eq:Vergleich_Channeluses_unit} holds: If the decoherence does not exceed a certain threshold, the pure state reconstruction requires less resources (channel uses) than the mixed state one.

In the second step, we can now compare these results to existing low-rank process tomography methods: 
\begin{enumerate}
    \item 
For a two-qubit process close to unitarity, the methods presented in Ref.~\cite{Kliesch2019quantum} and Ref.~\cite{Shabani2011PRL} require between $M \sim 50$ and $M \sim 150$ channel uses, 
whereas 
Eq.~\eqref{eq:channel uses pure unit} implies between $M = 8^2 = 64$ (for $r \approx 1$) and $M = 2^N 8^N = 256$ (for $r \approx \sqrt{d}$) channel uses for pure state reconstruction. 
\item The reconstruction of a 3-qubit Toffoli gate under weak noise requires between $M \sim 320$ and $M \sim 1000$ channel uses depending on the strength of the noise~\cite{Kliesch2019quantum}. 
In comparison, the scaling of our pure-state reconstruction of a three-qubit gate under weak noise requires $M = 8^3 = 512$ channel uses.  
\end{enumerate}
These two examples show that our methods are competitive in terms of the required resources for small systems.

For larger systems, we consider the 7-qubit channel tomography of Ref.~\cite{Surawy2022quantum} reporting $M_R=10^9$ channel uses, compared to our value of $M=16^7$, and a unitary 10-qubit channel tomography~\cite{torlai2020arxiv} for which $M = 10^{12}$ channel uses are used compared to $M = 16^{10}$ channel uses with our method.   We thus find that the protocols introduced here  are competitive with state-of-the-art approaches for low-rank quantum processes.


\section{Conclusions}\label{sec:conclusions}

We have presented two methods to approximate the unitary part of a non-unitary evolution using a set of $d+1$ pure, resp. two mixed, input states, where $d$ is the dimension of Hilbert space. 
These states were designed to collect all information relevant to certify a quantum channel, i.e. to compare the realized quantum evolution to a desired unitary~\cite{reich2013PRA}. The present work has extended this approach to the reconstruction of an $unknown$ unitary, providing explicit formulae that require only knowledge of the output states.
We have shown that reconstruction works well for no or weak decoherence. As expected, it struggles for strong decoherence where degeneracies in the image of the dynamical map hamper the meaningful identification of a specific unitary.  We find that reconstruction using pure input states offers a good compromise between accuracy and resource cost \correction{\footnote{The accuracy of all three methods depends only on the imperfection of the protocols (measurement errors, SPAM error etc) as well as the decoherence strength of the noise.}}. If accuracy is paramount, the Choi reconstruction provides the best results \correction{because the total amount of available information is greater than for the pure- and mixed state reconstruction}. \correction{Conversely, it also requires substantially more channel uses which may inhibit its practical use for increasing system sizes.} 
When comparing to compressed sensing methods~\cite{Gross2010arXiv,Flammia_2012,roth2018PRL}, in the regime of weak decoherence, the minimum number of channel uses is decreased from $M=\mathcal{O}(16^N)$ to $M=\mathcal{O}(8^N)$, where $N$ is the number of qubits, for the pure-state reconstruction. 
Moreover, comparing to compressed sensing methods exploiting the low rank of the quantum channel under weak noise~\cite{Kliesch2019quantum,Shabani2011PRL,Surawy2022quantum,torlai2020arxiv}, we find the resource scaling of our protocols to be competitive for both small and large $N$.
Our approach differs from these tomographic approaches which rely on a least-squares fit over sets of quantum channel~\cite{roth2018PRL}. Although this precludes to obtain analytical bounds using techniques from, e.g., compressed sensing, the lack of a numerical optimization removes a point of failure and reduces computational effort.

The most obvious application of the reconstruction methods presented here is in channel characterization~\cite{blumekohout2025arXiv,rosy2025detecting,morgillo2025learning}, 
quantum gate benchmarking~\cite{blumekohout2025arXiv} and possibly quantum error mitigation~\cite{Cai2023RMP-QuErrorMitigation}. 
\correction{For the latter, all coherent errors are readily mitigated by using the reconstructed unitary instead of the target unitary in circuit transpilation.}
Moreover, the approximate unitaries that the reconstruction provides, will allow for estimating the local invariants of a non-unitary two-qubit quantum channel. Alternatively, once could also employ multiparameter quantum estimation for extracting the local invariants. 
This allows for identifying the local invariants of a two-qubit unitary with a single input state, assuming the local parts are given by identity \cite{fazio2025arXiv}.
Identifying the local invariants of a not fully unitary evolution is relevant when extending the optimal control framework targeting a local equivalence class \cite{Mueller2011PRA} or an arbitrary perfect entangler \cite{Watts2015PRA,Goerz2015PRA} to open quantum system dynamics.


\section*{Acknowledgments}

Financial support from the German Federal Ministry for Education and Research (BMBF), Project No. 13N15929 QCStack, is gratefully acknowledged.

\bibliography{main}%


\appendix
\section{Proof for Unitary Reconstruction via Mixed Input States}\label{sec:appendix A}

Since $\mathcal{D}_{U}\left(\rho_{B}\right)$ originates from a unitary
transformation from $\rho_{B}$ the spectrum of $\rho_{B}$ and $\mathcal{D}_{U}\left(\rho_{B}\right)$
is identical. Because the spectrum is non-degenerate we can decompose
$\mathcal{D}_{U}\left(\rho_{B}\right)$ and $\rho_{B}$ in the following
way,
\begin{subequations}
\begin{eqnarray}
\mathcal{D}_{U}\left(\rho_{B}\right) & = & \sum_{i}\lambda_{i}{Q}_{i}\,,\\
\rho_{B} & = & \sum_{i}\lambda_{i}P_{i}\,,
\end{eqnarray}
\end{subequations}
where the $P_{i}/{Q}_{i}$ are unique, one-dimensional
projectors. Note that the ordering of the projector set $P_i$ and $Q_i$
is fixed by the ordering of the non-degenerate spectrum formed by the
eigenvalues $\lambda_i$.
Using these projectors we can write
\begin{equation}
\mathcal{D}_{U}\left(\rho_{B}\right)=\sum_{i}\lambda_{i}\tilde{P}_{i}=\sum_{i}\lambda_{i}UP_{i}U^{\dagger}=U\rho_{B}U^{\dagger}\,.
\end{equation}
Any unitary transformation $U$ applied to one-dimensional projectors leads
to a new set of one-dimensional projectors $UP_{i}U^{\dagger}=P_{i}^{\left(U\right)}$,
consequently
\begin{equation}
\sum_{i}\lambda_{i}Q_{i}=\sum_{i}\lambda_{i}P_{i}^{\left(U\right)}\,.
\end{equation}
Multiplication with $Q_{j}$ from the right side leads to
\[
\lambda_{j}Q_{j}=\sum_{i}\lambda_{i}P_{i}^{\left(U\right)}Q_{j}\,.
\]
Multiplication with $P_{k}^{\left(U\right)}$ from the left side leads
to
\begin{eqnarray*}
\lambda_{j}P_{k}^{\left(U\right)}Q_{j} & = & \lambda_{k}P_{k}^{\left(U\right)}Q_{j}\\
\Longleftrightarrow\left(\lambda_{j}-\lambda_{k}\right)P_{k}^{\left(U\right)}Q_{j} & = & 0\,.
\end{eqnarray*}
Due to non-degeneracy of the spectrum, i.e.~$\forall j\neq k:\,\lambda_{j}\neq\lambda_{k}$, it follows that
\begin{equation*}
\forall j\neq k:\ P_{k}^{\left(U\right)}Q_{j}=0\,,
\end{equation*}
whence follows (since the projectors were all one-dimensional)
\begin{equation*}
P_{j}^{\left(U\right)}=c_{j}Q_{j}\,,
\end{equation*}
with $c_{j}\in\mathbb{C}$. But since both $P_j$ and $Q_j$ are projectors we may
conclude that $\forall j:\,c_{j}=1$, i.e.
\begin{equation}
\forall j:\ Q_{j}=UP_{j}U^{\dagger}\,.\label{eq:projector mapping}
\end{equation}
This key relation allows for reconstruction of the unitary $U$ up to relative phases.
This can be seen as follows: Let $\ket{\psi_{j}}$
be a vector with \textbf{$Q_{j}\ket{\psi}_{j}=\ket{\psi}_{j}$}.
Then, by multiplication with $\bra{\psi_{k}}$ from the left
and $\ket{\psi_{k}}$ from the right, it follows from Eq.~\eqref{eq:projector mapping}
that
\begin{eqnarray}
\delta_{jk}&=&\braket{\psi_{k}|U|j}\braket{j|U^{\dagger}|\psi_{k}}\\
           &=&\left|\braket{\psi_{k}|U|j}\right|^{2}\,,\label{eq:more mappings}
\end{eqnarray}
with $\ket{j}$ given by $P_{j}\ket{j}=\ket{j}$.
As a result, $U$ must map $\ket{j}$ to a vector along $\ket{\psi_{j}}$
or else not all components for $j\neq k$ would vanish in Eq.~\eqref{eq:more mappings}.
Since $U$ is a isometry, i.e.~it preserves norms, we may conclude
for $U$ that
\begin{equation*}
\forall k:\ U\ket{k}=e^{i\varphi_{k}}\ket{\psi_{k}}\,,
\end{equation*}
for a set $\left\{ \varphi_{k}\right\} $ of yet to be determined
phases. Since we know the action of $U$ on an orthonormal basis we
may consequently write
\begin{equation}
U=\sum_{k}e^{i\varphi_{k}}\ket{\psi_{k}}\bra{k}\,.\label{eq:unitary_after_basefix}
\end{equation}
Note that while the vectors $\ket{k}$ and $\ket{\psi_k}$
are only determined up to a global phase, since they are just defined
as an eigenvector of a one-dimensional projector to eigenvalue $1$,
these phases can be absorbed in the $e^{i\varphi_{k}}$.

Finally, the phases $e^{i\varphi_{k}}$ in Eq.~\eqref{eq:unitary_after_basefix}
can be determined by applying the dynamical map $\mathcal{D}_{U}$
to the matrix $\rho_{P}$. The corresponding image is given by 
\begin{eqnarray*}
\mathcal{D}_{U}\left(\rho_{P}\right) & = & \frac{1}{d}\sum_{kl}\sum_{ij}e^{i\left(\varphi_{k}-\varphi_{l}\right)}\ket{\psi_{k}}\braket{k|i}\braket{j|l}\bra{\psi_l}\\
 & = & \frac{1}{d} \sum_{ij}e^{i\left(\varphi_{i}-\varphi_{j}\right)}\ket{\psi_i}\bra{\psi_j}
\end{eqnarray*}
By freedom of the global phase of the unitary
we may set $\varphi_{1}=0$ and it follows $\forall k=2,\dots,d$
that indeed
\begin{eqnarray*}
\Braket{\psi_{k}|\mathcal{D}_{U}\left(\rho_{P}\right)|\psi_{1}} & = & \frac{1}{d} e^{i\varphi_{k}}\nonumber \\
\Longrightarrow \varphi_{k} & = & \arg\left[d\braket{\psi_{k}|\mathcal{D}_{U}\left(\rho_{P}\right)|\psi_{1}}\right]\,.
\end{eqnarray*}

\section{CNOT-like gates for multiple qubits}\label{sec:appendix B}

\begin{figure}[hbt!]
     \centering
    \includegraphics[width=0.45\textwidth]{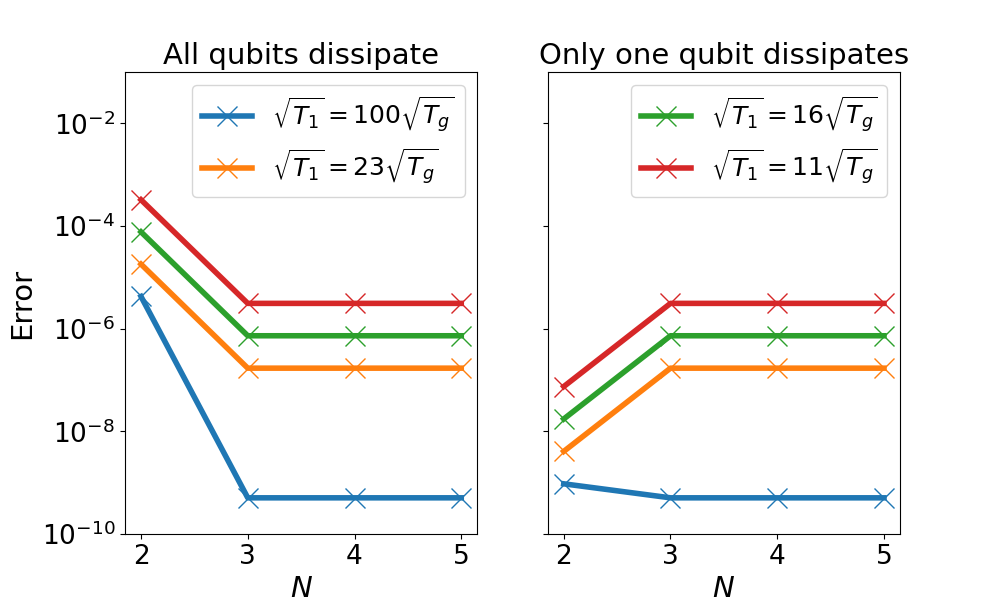}
           \caption{Behavior of reconstruction error $1-F_G(U_0,U_{RC})$ of the pure-state reconstruction method of CNOT-like gates for four designated $T_1$ times as a function of the number of qubits $N$. }
         \label{fig:Multiple qubits CNOT pure}
\end{figure}

\begin{figure}[hbt!]
     \centering
    \includegraphics[width=0.45\textwidth]{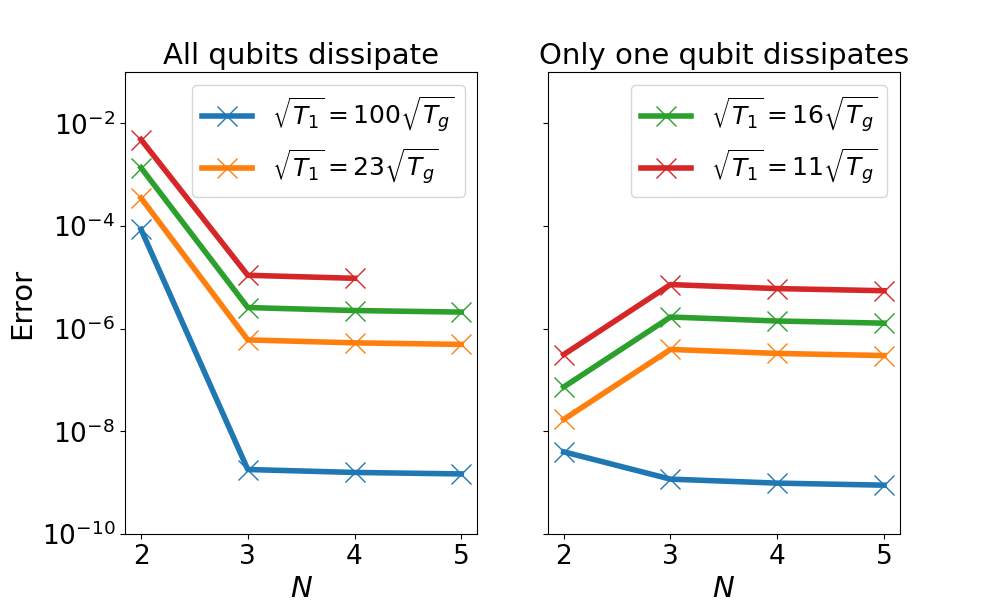}
           \caption{Behavior of reconstruction error $1-F_G(U_0,U_{RC})$ of the mixed-state reconstruction method of CNOT-like gates for four designated $T_1$ times as a function of the number of qubits $N$. }
         \label{fig:Multiple qubits CNOT mixed}
\end{figure}

For $N=2$, we consider the CNOT gate where the first (second) qubit is the control (target) qubit. For multiple qubits, i.e. $N \in \{3,4,5\}$, we assume that the extra qubits (as compared to the $N=2$ case) are additional control qubits while the target qubit remains the second one. We also assume that only a single qubit, namely the target qubit dissipates, so that the absolute dissipation is the same for every system size. Figures \ref{fig:Multiple qubits CNOT pure} and \ref{fig:Multiple qubits CNOT mixed} show the reconstruction error of the pure-state and mixed-state reconstruction methods respectively. They demonstrate that both methods can be used to reconstruct CNOT-like gates for multiple qubits, as indicated by the continuous reconstruction error graphs. However, in the left panel of Fig.~\ref{fig:Multiple qubits CNOT mixed} we notice that for the mixed-state reconstruction there are degeneracies in the image (see red graph), that do not appear in the pure-state reconstruction, cf. left panel of Fig.~\ref{fig:Multiple qubits CNOT pure}. When comparing the reconstruction errors of both methods, we again notice that the pure-state reconstruction performs better (compare position of the horizontal dotted blue line on the y-axis for the pure-state and mixed-state reconstruction).

\end{document}